\documentclass[prl,twocolumn,showpacs,superscriptaddress]{revtex4}
\usepackage{amssymb,color}
\usepackage[dvips]{graphicx}
\usepackage{verbatim}

\begin{document}

\title{Integrable two-channel $p_x+ip_y$-wave superfluid model}
\author{S.~Lerma~H.}
\affiliation{
  Departamento de F\'{i}sica, Universidad Veracruzana, Xalapa, 91000, Veracruz, Mexico
  }
\author{S. M. A. Rombouts}
\affiliation{
  Instituto de Estructura de la Materia,
  C.S.I.C.,
  Serrano 123, E-28006 Madrid, Spain}
\author{J. Dukelsky}
\affiliation{
  Instituto de Estructura de la Materia,
  C.S.I.C.,
  Serrano 123, E-28006 Madrid, Spain}
\author{G. Ortiz}
\affiliation{
  Department of Physics,
  Indiana University,
  Bloomington IN 47405, USA}

\begin{abstract}
We present a new two-channel integrable model describing a system of
spinless fermions interacting through a $p$-wave Feshbach resonance.
Unlike the BCS-BEC crossover of the $s$-wave case, the $p$-wave model
has a third order quantum phase transition. The critical point coincides
with the deconfinement of a single molecule within a BEC of bound
dipolar molecules. The exact many-body wavefunction provides a unique
perspective of the quantum critical region suggesting that the size of
the condensate wavefunction, that diverges logarithmically with the
chemical potential, could be used as an experimental indicator of the
phase transition.
\end{abstract}

\maketitle

In recent years $p$-wave paired superfluids have attracted a lot of
attention, in part  due to their exotic properties  \cite{Gura1}. Of
particular interest is the chiral two-dimensional ($2D$) $p_x+ip_y$
superfluid of  spinless fermions, that supports a topological phase with
zero energy Majorana modes \cite{Read}. The latter are  theorized to
serve as a basic element for a topological quantum computer
\cite{Nayak}.  That exotic superfluid state might be realized in the A1
phase of $^{3}$He \cite{Volo}, in the layered perovskite oxide
Sr$_2$RuO$_4$ \cite{Sarma}, and in the Pfaffian quantum Hall state at
$\nu=5/2$ filling \cite{Moore}.
Most promising is its realization in a cold gas of fermionic atoms in a
single hyperfine state. Indeed, $p$-wave Feshbach resonances have been
observed and studied in $^6$Li and $^{40}$K \cite{Fesch} with the
potential to manipulate the system from the weak (BCS) to the strong
(BEC) pairing regime.  However, these gases revealed to be unstable due
to atom-molecule relaxation processes in which the molecule decays to a
deep bound state while the atom escapes with excess energy \cite{Levin}.
Other atomic and molecular gases are now in consideration, as well as
different mechanisms to suppress relaxation.

{}From a theoretical standpoint, despite great efforts to describe these
systems, a complete understanding of the BCS-BEC transition and the
corresponding phase diagram is still missing. Recently, by means of an
exactly solvable $p_x+ip_y$ pairing model \cite{Sierra, Dunning2010} it
was shown that the quantum phase transition (QPT) taking place from weak
pairing to strong pairing can be understood as  the deconfinement of
bound Cooper pairs \cite{Romb}. In this Letter we introduce an exactly
solvable two-channel $p_x+ip_y$ pairing model  with a Feshbach
resonance. Our model is exactly solvable in arbitrary dimensions,
although we will  concentrate on its $2D$ realization. We propose a way
to experimentally detect the so-called {\it topological} QPT
\cite{volovik} from weak to strong pairing, by measuring a
density-density correlation function. This together with the analysis of
the size of a Cooper pair in terms of the exact solution allows the 
characterization of the transition as one of a confinement-deconfinement
type without Landau order parameter. Moreover, the transition is shown
to be third order in the Ehrenfest classification.  Another way to
theoretically detect that QPT is by analyzing the behavior of the
quantum fidelity $z$ of the ground state wavefunction. Interestingly,
the second order derivative of $\ln |z|$ displays a logarithmic
singularity at the transition point confirming its third order
character.
%

Consider the $2D$ two-channel $p_x+ip_y$-wave model
\begin{eqnarray} \label{Ham}
H&=&\sum_{k,k_{x}>0}\frac{|k|^{2}}{2}\left( \hat{n}_k +\hat{n}_{-k}
\right) + H_b \\
&-&g\!\!\!\sum_{k,k_{x}>0}\left[ \left(
k_{x}+ik_{y}\right) b \, c_{k}^{\dagger }c_{-k}^{\dagger }+\left(
k_{x}-ik_{y}\right) b^{\dagger }c_{-k}c_{k}\right] ,\nonumber
\end{eqnarray}
where $c^\dagger_k$ creates a fermion in mode $k=(k_x,k_y)$,
$\hat{n}_k=c_{k}^{\dagger}c^{\;}_{k}$, $b^\dagger$ is a bosonic creation
operator, and  $H_b=\delta \, b^{\dagger }b+g^{2}\, b^{\dagger
}b^{\dagger }bb$. Our next goal is to show that this model is a
particular realization of a family of exactly-solvable atom-molecule
Hamiltonians of physical relevance in the context of cold atom physics.

We start our demonstration by recalling the integrals of motion of the
hyperbolic Richardson-Gaudin model \cite{Duke1}, which can be
generically written
as \cite{NuclPhysB}
\begin{eqnarray}
R_{i}&&=S_{i}^{z}- \label{Rin}\\
&&2\gamma \sum_{j\not=i} \left[ \frac{\sqrt{\eta _{i}\eta _{j}} }{\eta
_{i}-\eta _{j}}\left ( S_{i}^{+}S_{j}^{-}+S_{i}^{-}S_{j}^{+}\right]) +
\frac{\eta _{i}+\eta _{j}}{\eta _{i}-\eta _{j}}S_{i}^{z}S_{j}^{z}\right]
\nonumber ,
\end{eqnarray}
where $S_{i}^{z}$, $S_{i}^{\pm}$, are the three generators of the
$SU(2)_{i}$ algebra of mode $i$, $i=0,\cdots,L$, with spin
representation  $s_{i}$ such that $\langle S_i^2\rangle=s_i(s_i+1)$. 
Therefore, the operators $R_{i}$ contain $L+1$ free
parameters $\eta_{i}$ plus the strength of the quadratic term $\gamma$.
The integrals of motion (\ref{Rin}) commute with the $z$ component of
the total spin component $S^{z}=\sum_{i=0}^L S_{i}^{z}=M-\sum_{i=
0}^Ls_{i}$.

We next  single out the copy $i=0$  and consider its 
large spin $s_0$ limit; eventually we are interested  in the limit $s_{0}\rightarrow \infty
$. The corresponding $SU(2)_{0}$ generators are bosonized by means of
the Holstein-Primakoff mapping
\begin{eqnarray}
\!\!\! S_{0}^{z}=b^{\dagger }b-s_{0}, \quad S_{0}^{+}=b^{\dagger }\sqrt{
2s_{0}-b^{\dagger }b},\quad S_{0}^{-}=(S_{0}^{+})^\dagger . \nonumber
\end{eqnarray}
In the spin-boson representation  the conservation of $S^z$ becomes
$b^{\dagger }b+\sum_{i=1}^L S_{i}^{z}=M-L_{c}/2$, 
where $L_c=2\sum_{i=1}^L s_{i}$.

Inserting the boson representation into the integrals of motion and
expanding them in terms of $1/s_{0},$ we arrive to the complete set
of integrals of motion describing a spin-boson model
\begin{eqnarray}
\mathcal{R}_{0} &=&H_b+\sum_{j}\eta _{j}S_{j}^{z}-g\sum_{j}\sqrt{\eta
 _{j}}\left( b^{\dagger }S_{j}^{-}+bS_{j}^{+}\right) \label{Int0}\\
\mathcal{R}_{i} &=&\left( \eta _{i}+\frac{\kappa g^{2}}{2}
\right)S_{i}^{z}-g^{2}S_{i}^{z}b^{\dagger }b-g\sqrt{ \eta _{i}}\left(
b^{\dagger }S_{i}^{-}+bS_{i}^{+}\right)   \nonumber \\
+&& \!\!\!\!\!\!\!\!\!\! g^{2}\sum_{j(\not=i)}\left[ \frac{\sqrt{\eta
_{i}\eta _{j}}}{\eta _{i}-\eta _{j}}\left(
S_{i}^{+}S_{j}^{-}+S_{i}^{-}S_{j}^{+}\right) +\frac{ \eta _{i}+\eta
_{j}}{\eta _{i}-\eta _{j}}S_{i}^{z}S_{j}^{z}\right] \nonumber ,
\end{eqnarray}
where we took advantage of the freedom to select the values of $\gamma$
and $\eta_0$: $\gamma =\frac{1}{2s_{0}}+\frac{\kappa }{4s_{0}^{2}}
{\hbox{    and  }}\eta _{0}=2g^{2}s_{0}$, so that finite integrals of
motion result in the limit $s_0\rightarrow \infty$. The detuning
parameter $\delta$ in $H_b$ is given by
$\delta=(L_c-2(M-1)-\kappa)g^2/2$, with $\kappa$ and $g^2$ free
parameters.

The corresponding eigenvalues \cite{NuclPhysB} in this limit are
\begin{eqnarray}
\mathit{r}_{0} &=&\sum_{\alpha }E_{\alpha }-\sum_{j}\eta _{j}s_{j}
\label{eigen}\\
\mathit{r}_{i} &=& s_i g^2\left(\sum_{j(\not=i)}s_{j}\frac{\eta
_{i}+\eta _{j}}{\eta _{i}-\eta _{j}}+\sum_{\alpha }\frac{E_{\alpha
}+\eta _{i}}{E_{\alpha }-\eta _{i}}- \frac{\eta
_{i}}{g^2}-\frac{\kappa}{2}\right ).  \nonumber
\end{eqnarray}
where the set of pair energies (pairons) $E_{\alpha}$ represents a
particular solution of the Richardson equations
\begin{equation}
\frac{1}{2g^{2}}+\sum_{i}\frac{s_{i}}{\eta _{i}-E_{\alpha
}}-\sum_{\alpha ^{\prime }(\not=\alpha )}\frac{1}{E_{\alpha ^{\prime
}}-E_{\alpha }}=\frac{ Q_{o}}{E_{\alpha }},
\label{Rich2}
\end{equation}
with $Q_o=(M-1)-L_c/2+(\delta/2 g^2)$. The exact eigenstates in turn
are given by
\begin{eqnarray}
\left\vert \Psi \right\rangle =\prod\limits_{\alpha =1}^{M}\left(
b^{\dagger }+g\sum_{i}\frac{\sqrt{\eta _{i}}}{\eta _{i}-E_{\alpha }}
\, S_{i}^+ \right) \left\vert 0, \nu\right\rangle ,
\label{WF}
\end{eqnarray}
where $\left\vert 0, \nu\right\rangle$ represents the boson vacuum
tensor and the state of seniority $\nu$, such that $S^z_i \left\vert\nu
\right\rangle = -s_i \left\vert\nu \right\rangle$ and $S^-_i
\left\vert\nu \right\rangle =0$ for all $i$.

This completes the derivation of the integrable spin-boson model defined
by the integrals of motion (\ref{Int0}), whose eigenvalues and common set of
eigenvectors are given by (\ref{eigen}) and (\ref{WF}) respectively,
expressed in terms of the pair energies $E_\alpha$ solutions of the
Richardson equations (\ref{Rich2}).

The connection between the spin-boson model above and the
$p_{x}+ip_{y}$  pairing model of Eq. (\ref{Ham}) is realized by the pair
representation of  the $SU(2)$ algebra $S_{k}^{z}=\frac{1}{2} \left(
\hat{n}_{k}+\hat{n}_{-k}-1\right) ,~S_{k}^{+}=\left( S_{k}^{-}\right)
^{\dagger }=(k_{x}+ik_{y}) c^\dagger_k c^\dagger_{-k}/ \left\vert
k\right\vert$, where now we consider the mode index to represent
momentum $k=(k_x, k_y)$. Now $M$ represents the total number
of fermionic pairs and bosons and $L_c$ is the maximum possible number 
of fermionic pairs. Inserting this representation in the integral
of motion $\mathcal{R}_{0}$ and defining $\eta_k=|k|^2$ we obtain the
Hamiltonian (\ref{Ham}), $H=\mathcal{R}_{0}$, thus showing that it is
exactly solvable. Indeed, any linear  combination of the integrals of
motion defines an exactly solvable  Hamiltonian.  In particular, the
Hamiltonian studied by Links \textit{et al.} \cite{links2011}, can be
obtained from the linear combination $ \widetilde{H}=2\sum_{i}\epsilon
_{i}\mathcal{R}_{i}$ with $2\epsilon _{i}=1/\eta _{i}$. If $\epsilon
_{k}=|k|^{2}/2$, with $0\leqslant k\leqslant k_{\sf cut}$, then the
parameters $\eta_{k}$ entering in the Richardson's equations (\ref{Rich2}) are
defined in the interval  $1/|k_{\sf cut}|^2=1/(2\omega)\leqslant \eta
_{k}\leqslant \infty $. Their exactly solvable model  does not display a
QPT between the weak and strong pairing phases, because, as stated in
Ref. \cite{Romb}, for the $p_{x}+ip_{y}$ model to display a non-analytic
behavior in the continuum limit it is required that one of the
parameters $\eta_k$ vanishes for a given mode (e.g., $k=0$ mode).

\begin{figure}
\includegraphics[angle=0,width=0.4\textwidth]{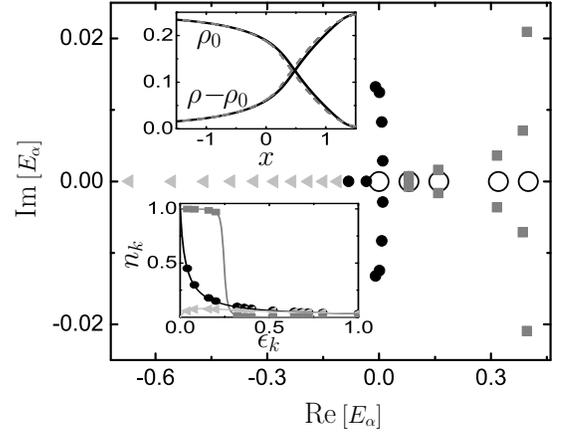}
\caption{Pairon distribution for  $x=-0.1$ (triangles), $x=0.37$ (solid
circles) and $x=1.7$ (squares) in a system of $M=10$ pairons, $L_c=40$
($\rho=1/4$),   $\lambda=g^2 L_c=1/2$, and $\omega=1$. The open circles
represent the five lowest level parameters $\eta_k$. The upper inset
shows the bosonic, $\rho_0$, and fermionic, $\rho-\rho_0$,  densities
for the exact solution (solid line) and the  BCS approximation (dashed
line). The  lower inset displays the exact momentum  distribution,
$n_k$, for the three cases and the BCS approximation in solid lines.}
\label{fig2}
\end{figure}

Coming back to the Eqs.  (\ref{Rich2}), in complete analogy  with Ref.
\cite{Romb}, we can recognize two special cases. One in which all the
pairons $E_\alpha$ converge to zero, corresponding to $2Q_o+1=M$, and a
second case where the first pairon converges  to zero, i.e., when
$2Q_o+1=0$. Defining $x=\delta/ (2 g^2 L_c)$, these two cases
correspond  to $2x_{\sf MR}=1-(M-1)/L_c$ and $ 2x_{\sf
cr}=1-(2M-1)/L_c$. The first case defines the so-called Moore-Read line.
We will show that the second case  signals an interesting third-order
QPT.

To get insight into the properties of the different phases we analyze
the behavior of the pairons in a finite system.  Figure  \ref{fig2}
displays the pairon distribution for $x<x_{\sf cr}$ (strong pairing),
$x_{\sf cr}<x< x_{\sf MR}$ and $x_{\sf MR}<x$ (weak pairing on both
sides of the Moore-Read line) for a system consisting of $M=10$ pairs
lying in a disk of  radius five units in an otherwise square lattice
with $L_c=40$ ($\rho\equiv  M/L_c=1/4$) and $\omega=1$. For $x=1.7$ the system has a
fraction of 6 Cooper pairs  (complex pairons) and 4 quasi-free pairs
states (almost real and positive pairons). Crossing the 
Moore-Read line, at $x=0.37$, the system is a  mixture of 8 Cooper pairs and 2 bound
pairs (real an negative pairons).  Well inside the strong pairing phase,
at $x=-0.1$, all pairs are bound.  Therefore, the weak pairing phase is
characterized by a mixture of free  fermions, Cooper pairs and bound
molecules, while the strong pairing  phase is a BEC of bound molecules.
The upper inset displays  the bosonic density, $\rho _{0}=\left\langle
b^{\dagger }b\right\rangle /L_c$, while the lower inset shows  the major
rearrangement that takes place in the momentum distribution  $n_k$ close
to $k=0$ between the weak and the strong pairing phases.

\begin{figure}[htb]
\includegraphics[angle=0,width=0.4\textwidth]{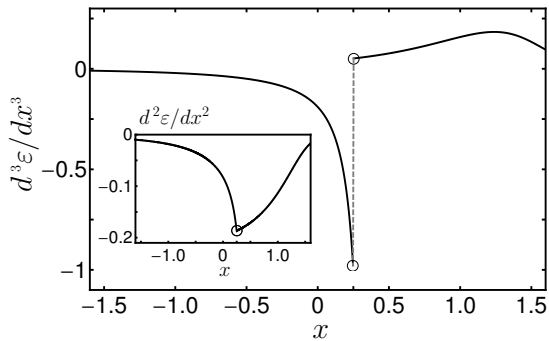}
\caption{Second (inset) and third-order derivatives of the energy
density $\varepsilon$ in the thermodynamic limit ($\rho=1/4, \lambda=1/2$, and $\omega=1$
leading to $x_{\sf cr}=1/4$).}
\label{fig3}
\end{figure}

Let us now consider the thermodynamic limit ($L_c\rightarrow\infty$ with
$\lambda=g^{2}L_c$ finite) of this two-channel $p$-wave model.
Following the same procedure as in \cite{Romb} we obtain a pair of
couple BCS-like equations for the unknowns $\rho_0$ and chemical
potential $\mu$


\begin{eqnarray}
&&\hspace*{-0.6cm} x+\rho_0-\frac{\mu}{\lambda}=\label{BCS1}\\
&&\frac{1}{2\omega}\left[f-\vert\mu\vert+\left(\mu-\lambda\rho_0\right)
\ln\left( \frac{\omega-\mu+\lambda \rho _{0} +f}{\lambda \rho _{0}-
\mu+\left\vert \mu \right\vert }\right)   \right] ,\nonumber\\
&&\hspace*{-0.6cm} \rho-\rho_0= \label{BCS2}\\
&&\frac{1}{2}-\frac{1}{2\omega} \left[f-\vert\mu\vert-\lambda\rho_0
\ln\left( \frac{\omega-\mu+\lambda \rho _{0} +f}{\lambda \rho _{0}-
\mu+\left\vert \mu \right\vert }\right) \right],\nonumber
\end{eqnarray}
where $f=\sqrt{2\lambda\omega\rho
_{0}+\left( \omega-\mu \right) ^{2}}~$.

The ground state energy density $\varepsilon=E/L_c$ is
\begin{eqnarray}
\varepsilon=\rho(\omega+\mu)+\lambda\rho_0 x-
\left(x-\frac{1}{2}+\rho\right)\frac{\omega
\lambda\rho_0}{\mu}-\frac{\mu +\left\vert \mu \right\vert}{2}.\nonumber
\end{eqnarray}

\begin{figure}[tb]
\includegraphics[angle=0,width=0.40\textwidth]{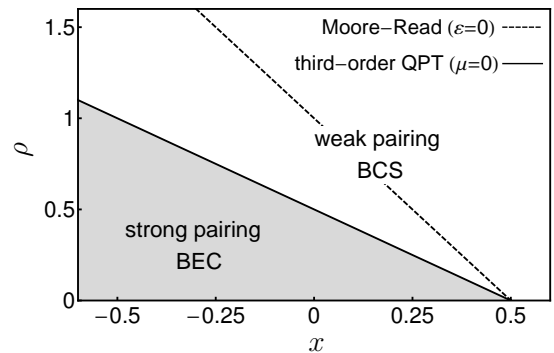}
\caption{Quantum phase diagram  in terms of $\rho$ and $x$, for a fixed
value of $\lambda$. Phase boundaries are insensitive to $\lambda$.}
\label{fig1}
\end{figure}

If we identify $\lambda \rho _{0}=\Delta^2_{\sf BCS}$ and $x=1/2g_{\sf
BCS}$, the energy density $\varepsilon$ coincides with the energy
density for the  one-channel $p_x+ip_y$ model \cite{Romb}. Analogously,
the right hand sides of Eqs. (\ref{BCS1}-\ref{BCS2}) coincide with the
corresponding ones in  \cite{Romb}.   However, their left hand sides are
different due to the existence of two independent parameters $x$ and
$\lambda$, or equivalently $\delta$ and $g$. Potential non-analyticities
at $\mu=0$  can be attributed to the presence of absolute value of $\mu$
terms.  Indeed, from Fig.  \ref{fig3} that shows the second- and
third-order derivatives of the energy density with respect to the
control parameter $x$, we conclude that the $p$-wave atom-molecule model
displays a third-order QPT at the critical value $x_{\sf
cr}=\frac{1}{2}-\rho$, coinciding with the limit in which all pairons
are real and negative except one that converges to zero, and $\mu=0$. The Moore-Read 
line corresponds to $x_{\sf MR}=(1-\rho)/2$ and $\mu=\lambda \rho_0/2$, 
leading to $\varepsilon=0$, coinciding with the limit in which all pairons 
converge to zero.  The resulting
quantum phase diagram depicted in Fig. \ref{fig1}, depends on the density $\rho$,  
and on the control
parameters $\lambda$ and $x$. However, the phase  boundaries are
independent of $\lambda$.  Thus, our
two-channel $p$-wave model extends the one-channel model  to:  a) $x<0$
due to the possibility of having  negative detunings $\delta$, b)
$\rho>1$ due to the mixture with a bosonic system, and c) include an
extra control parameter $\lambda$.

\begin{figure}[htb]
\includegraphics[angle=0,width=0.40\textwidth]{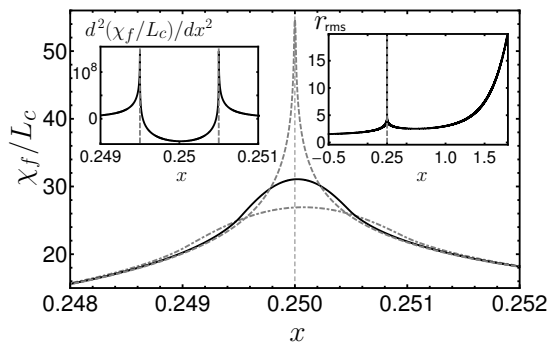}
\caption{ $\chi_f/L_c$ (see text) as a function of $x$,
$\rho=1/4$,  $\lambda=1/2$, $\omega=1$ for $\Delta x=10^{-3}$ (dot-dashed) , $5\times
10^{-4}$ (solid), and  $10^ {-5}$  (dashed). At $x_{\sf cr}=1/4$,
as $\Delta x\rightarrow 0$, it diverges logarithmically. At $x=x_{\sf
cr}\pm\Delta x$, and finite $\Delta x=5\times 10^{-4}$, the second-order
derivative  of $\chi$ develops another logarithmic divergence in terms
of $\mu$.  The root mean square of the Cooper pair size, $r_{\sf rms}$
is shown in the right inset.}
\label{fig4}
\end{figure}

Since there is no Landau order parameter characterizing this third-order
transition, one would like to devise a way to experimentally detect it.
In Ref. \cite{Romb} the root mean square of the condensate wave function
was proposed  as a possible indicator of that QPT
\begin{equation}
r_{\sf rms}^2=\frac{\int\vert\nabla\phi(k)\vert^2 d^2k}{\int
\vert\phi(k)\vert^2 d^2k},
\label{r2}
\end{equation}
with $\phi(k)\propto   \frac{k_x+i k_y}{\sqrt{(|k|^2-a)(|k|^2-b)}} $, where
$a$ and $b$ are the roots of the polynomial
$\xi_k^2=(\epsilon_k-\mu)^2+2 \lambda \rho_0 \epsilon_k \equiv
(\epsilon_k-a/2)(\epsilon_k-b/2)$. 
The integral (\ref{r2}) can be obtained  analytically.
As can be seen in the right inset of Fig. \ref{fig4}, $r_{\sf rms}$
diverges logarithmically at the QPT point as $r_{\sf rms}^2\rightarrow
-\frac{\ln \vert \mu/\sqrt{2\omega\lambda\rho_0} \vert}{2\lambda\rho_0\ln
\left(1+\omega/2\lambda\rho_0\right)}$. Making contact with the exact wave
function (\ref{WF}) we can immediately associate this divergence to the
confinement of the last Cooper pair or, coming from strong pairing, to
the deconfinement of a single bound pair within the BEC molecular wave
function. Away from the QPT point, $r_{\sf rms}$ diverges again in the
extreme weak coupling limit, $x\rightarrow \infty$, in which all pairs are 
deconfined. The condensate wave function can be related to the
density-density correlation function  which in coordinate space can be
expressed as
$$
|\phi(r-r')|^2= \langle (\hat{n}_r- \langle \hat{n}_r \rangle)
(\hat{n}_{r'}-\langle \hat{n}_{r'}\rangle ) \rangle+|F(r-r')|^2 ,\nonumber
$$
where $\phi(r-r')$ is the condensate wave function in coordinate
space,  $\hat{n}_r=c^{\dagger}_r c_r$, $\langle  \hat{n}_r \hat{n}_{r'}
\rangle$ is the density-density correlation function, and $F(r-r')$ is
the  Fourier transform of the momentum density $\langle \hat{n}_k\rangle$.
In trapped cold atomic gases,  the condensate wave function can be
obtained from measurements of the density-density  correlations using
quantum noise interferometry and the momentum distribution from
time-of-flight measurements after opening the trap. Therefore, the root
mean square size of the condensate wave function constitutes a unique
indicator of the QPT which can be experimentally accessed.

The quantum fidelity $z[x,\Delta x]=\langle\Psi(x-\Delta x)|\Psi(x+\Delta x)
\rangle$ 
 can also be used as an indicator of QPTs  \cite{ours1}. The 
quantity $ \chi_f= -2\ln \vert z[x,\Delta x] \vert/\Delta x^2$, which in the limit $\Delta x\rightarrow 0$ is  the so-called fidelity susceptibilty,  has an
essentially different behavior depending on the ground-state overlaps
considered. If the overlap is taken between two states belonging to the
same phase ($\vert x-x_{\sf cr}\vert>\Delta x$), 
$\chi_f/L_c$ tends to a value which remains  finite  in the limit
$\Delta x\rightarrow 0$, whereas if  the overlap is between two states
in different phases ($\vert x-x_{\sf cr}\vert\leq\Delta x$) a 
$\ln\Delta x$ dependence   appears. More precisely,  $ \left.
(\chi_f/L_c\right)\vert_{x_{\sf cr}}\approx    \left. -(d\mu/dx)^2 \ln
(\Delta x) /(\omega\lambda \rho_{0})\right\vert_{x_{\sf cr}}$.
Clearly, when $\Delta x\rightarrow 0$  the fidelity  susceptibility
develops a logarithmic divergence as a function of $(x-x_{\sf cr})$ similar to that
obtained in the one-channel $p_x+ i p_y$ model \cite{Dunning2010}. 

However, even for $\Delta x$ finite, 
$\chi_f/L_c$ shows 
a non-analytic behavior. When one of the ground-states in $z[x,\Delta x]$ is
taken close to  the  transition point  ($x\approx x_{\sf cr}\pm\Delta
x$), 
$\chi_f/L_c$ can be written as  $\left.
(\chi_f/L_c)\right\vert_{x\approx x_{\sf cr}\pm\Delta x} =-
\mu_\mp^2  \ln \vert\mu_\mp \vert/(2\omega\lambda\rho_0 \Delta x^ 2)+\mathcal{O}(\mu_\mp^0)$, 
with $\mu_\pm=\mu (x \pm \Delta x )$.  At  $x=x_{\sf cr}\pm\Delta x$,
where $\mu_\mp=\mu(x_{\sf cr})=0$, the previous expression is 
continuous, but its second-order derivative  diverges logarithmically,
as shown in Fig. \ref{fig4}.   These results should be contrasted with 
the results obtained for  an  Ising  chain of length $L$ 
\cite{damski},  where  $\left. (\chi_I/L)\right\vert_{x_{\sf cr}}
\approx 1/|\Delta x|$, i.e. power law,  and   a logarithmic  divergence at $x=x_{\sf
cr}\pm\Delta x$  appears in the  first-order
derivative of $\chi_I/L$.

In conclusion, we presented a new exactly solvable spin-boson model 
which has as particular realization the two-channel $p_x+ip_y$-wave superfluid.
We showed that the model has a third order QPT that can be accessed 
experimentally by measuring the density-density correlation function. 
The analysis in terms of fidelity provides further evidence of the non-Landau 
character of the QPT, and its logarithmic singularity indicates that it is of a 
confinement-deconfinement type, as suggested by the exact wave function.


We acknowledge support from a Marie Curie Action of the European
Community Project No. 220335, the Spanish Ministry for Science and
Innovation Project No. FIS2009-07277, and  the Mexican Secretariat
of Public Education Project PROMEP 103.5/09/4482.

\end{document}